\documentclass[aps,twocolumn,showpacs]{revtex4}
\usepackage{graphicx}
\usepackage{epsfig}

\begin{document}
\def\f{\bar{F}}
\def\R{{\bf R}}
\def\Itens{\mbox{\sffamily\bfseries I}}

\title{Collective stochastic resonance in shear-induced melting of
sliding bilayers}
\author{Moumita Das$^{1}$\cite{a}, G. Ananthakrishna$^{2,1}$\cite{b}
and Sriram Ramaswamy${}^{1}$\cite{c}}
\affiliation{${}^1$Centre for Condensed Matter Theory, Department of Physics,
Indian Institute of Science, Bangalore 560 012, INDIA\\
${}^2$ Materials Research Centre, Indian Institute of Science, Bangalore 560 012 INDIA}
\date{\today}
\begin{abstract}
The far-from-equilibrium dynamics of two crystalline two-dimensional 
monolayers driven past each other is studied using Brownian dynamics 
simulations. While at very high and low driving rates the layers slide 
past one another retaining their crystalline order, for intermediate range 
of drives the system alternates irregularly between the crystalline and 
fluid-like phases. A dynamical phase diagram in the space of interlayer 
coupling and drive is obtained. A qualitative understanding of this stochastic 
alternation between the liquid-like and crystalline phases is proposed in 
terms of a reduced model within which it can be understood as a stochastic 
resonance for the dynamics of collective order parameter variables. This 
remarkable example of stochastic resonance in a spatially extended system 
should be seen in experiments which we propose in the paper.
\end{abstract}

\maketitle

\section{INTRODUCTION AND RESULTS}
\subsection{Background}
The shear flow of a solid is one of the most important and widely
studied \cite{prbk,mrbk,balentsetal,giamarchi} nonequilibrium phenomena in 
materials science, with relevance to such practical problems as the yielding
of materials, solid friction and even the mechanical properties of
the earth's crust. Such flow takes place when solids are subjected
to stresses which range from a few percent of the shear modulus
to, in some cases, a value of the order of the shear modulus
itself. It is particularly convenient to study such phenomena
using very soft solids, where the desired stress to modulus ratio
is easily achieved. Indeed, such studies open up new regimes in
the physics of driven systems. A variety of such unconventional,
ultra-soft solids have been studied, including packings of
multilamellar vesicles \cite{salmon}, vortex lattices in type II
superconductors \cite{fll}, and crystalline arrays,
electrostatically or sterically stabilized, of colloidal particles
in aqueous suspensions \cite{collsusp}. Experiments on suspensions
of interacting colloidal particles under shear are of particular
interest to us here, for the rich range of interesting phenomena
they reveal, including the shear-induced distortion of the static
structure factor in the fluid state, and stick-slip dynamics
\cite{stickslip}, hysteresis \cite{hystcolloid} and shear induced
melting \cite{ackerson,chaikin}, in the crystalline state. It is
likely that the properties of sheared crystals, as observed in
macroscopic three-dimensional scattering studies or in time- or
frequency-domain mechanical measurements, are the average result
of many intermittent, spatially inhomogeneous internal events. 
Accordingly, this paper focuses on such events, at the
level of the relative motion of an adjacent pair of layers, since
we believe that knowledge of these events will greatly aid our
understanding of the mechanisms underlying phenomena such as
shear-melting. We emphasize at the outset, to avert any confusion
on this score, that the phenomena which our study uncovers, and
which we discuss in detail below, are quite distinct from the 
well-known stick-slip effect in atomically thin fluid films subjected to 
shear \cite{thompson,homola,drummond,braun}.  

One popular approach to the study of sheared solids has been to
consider an ordered layer (the adsorbate), dragged over a fixed,
rigid periodic potential (the substrate), the latter representing
an adjacent layer \cite{frenkel,pr,gr}. This description is clearly 
limited in its applicability since it rules out deformation of the substrate,
although it is a reasonable starting point for experimental
situations in which the overlayer is much softer than the
substrate. It is natural, and more general, to ask instead what
happens when both adsorbate and substrate are dynamical, and
organize themselves into various structures, depending on
interaction strengths, temperature and driving force, and it is in
this spirit that our model is formulated. The case where both
layers are comparably deformable, in particular, is clearly of
relevance to sheared crystals. In all cases, each layer confronts
a periodic potential produced by the other layer, but both
amplitude and phase of this periodic potential are dynamical and
change as a result of interactions, noise and driving force,
giving rise to some remarkable collective effects, reported
briefly earlier \cite{msg} and discussed in detail in this paper.

Although the primary motivation for this paper was the problem of
sheared colloidal crystals, there are two other classes of problem
to which our study has a natural connection. One is the phenomenon
of lane formation in counter-driven interacting particles
\cite{lowen}, the other is the equilibrium modulated-liquid to
solid transition of interacting particles in an external periodic
potential. We will touch upon the relation of these problems to
our work later in this paper.

\subsection{Summary of models and results}

We report two detailed studies in this paper: first, a Brownian
dynamics simulations of a many-particle model \cite{msg},
henceforth referred to as the particle model, and second, a
reduced model, introduced to get insight into the results of the
particle model,  consisting of just two degrees of freedom
\cite{rlsrsmelt}, an order parameter amplitude and a strain field.
The particle model consists of two species of particles, A and B,
driven by a force F, of constant magnitude, in opposite directions
say, along $+x$ and $-x$ respectively (Fig.~\ref{model}). The
$V_{AA}$ and $V_{BB}$ interactions are identical. The $V_{AB}$
interaction has the same form but is smaller by a factor
$\epsilon$. This factor, in a phenomenological way, incorporates
the physics of the third direction (see below). All pairwise
interactions are of screened Coulomb form, with the screening
parameter so chosen that, when $F=0$, each species in the absence of 
the other settles down in a macroscopically
ordered triangular lattice configuration. The dynamics of the
system is modeled by the overdamped Langevin equation for
relatively sheared sets of particles and is monitored for
different F and $\epsilon$. With the inter-layer coupling strength
$\epsilon$ held constant, on increasing the drive, we observe an
interesting sequence of non-equilibrium states, namely, a sliding
crystalline ordered state (Fig.~\ref{slidingorder}), a sliding melt-freeze state
(characterized by alternate states of  order and disorder in
time), followed again by a sliding ordered state.
 In the intermediate ``melt-freeze'' regime, for fixed drive, the residence 
time of the system in the ordered state decreases and that of the disordered
state increases as a function of 
$\epsilon$ (Figs.~\ref{skmedium},~\ref{sklow},~\ref{skhigh}). 
The allowed nonequilibrium states  are best understood in
terms of a dynamic phase diagram of these states. We present such 
a dynamical nonequilibrium phase diagram (Fig.~\ref{phase1}) demarcating
the three regimes (i) lower smooth sliding, (ii) alternating
melt-freeze state, and (iii) upper smooth sliding state. 
The melt-freeze alternations are most pronounced in a window of driving 
force F and inter-layer coupling $\epsilon$ values. These melt-freeze
cycles are strongly reminiscent of the time series of a system
undergoing stochastic resonance \cite{benzi,nicolis,stochres,wisen} 
and, to explore this aspect in more detail following ref.\cite{rlsrsmelt}, 
we introduce the reduced model. Using the reduced model, we study the time
evolution of the system using coupled time dependent Ginzburg
Landau equations for the order parameter and strain fields, as a
function of a coupling parameter ($\alpha$) entering the model
equations and a drive ($\Omega$) analogous to $\epsilon$ and $F$ 
respectively. For a certain range of values of
$\alpha$, keeping $\alpha$ fixed, as a function of the drive
parameter $\Omega$, we observe three regimes (Fig.~\ref{diffdrivetime}
and Fig.~\ref{diffdriveprob})
- a crystalline state ( nonzero order parameter value), a bistable
regime where the system  alternates between the crystalline and
liquid state (order parameter values being zero), followed again
by a crystalline state. Keeping $\Omega$ fixed at an optimum
value, we find that the ratio of the average lifetime of the
crystalline state to that of the the liquid state in the
intermediate regime of bistability decreases as $\alpha$ is
increased (Fig.~\ref{diffcoupltime} and Fig.~\ref{diffcouplprob}). 
These observations are remarkably similar to the
phenomenon observed in the particle model and indeed the phase 
diagrams of the two models (Fig.~\ref{phase1} and Fig.~\ref{phase2})
correspond surprisingly well. 
Further, the reduced model exhibits a maximum in the signal to
noise ratio at optimum values of the noise intensity (Fig.~\ref{snr}), thereby
making the connection to stochastic resonance concrete 
~\cite{benzi,nicolis,stochres,wisen}.

The paper is organized as follows. The Brownian dynamics
simulations of the particle model are described in Section II A 
and the results discussed in detail in section II B. 
This is followed by physical arguments in support of
the behaviors observed. The reduced model \cite{rlsrsmelt} is
introduced in Section III A and its results discussed in III B. 
Finally in Section IV we provide a discussion of
our results, suggest how our observations  may be verified
experimentally and outline directions of future research.

\section{BROWNIAN DYNAMICS SIMULATIONS OF TWO ADJACENT MONOLAYERS }

\subsection{Particle Model}

\begin{figure}
\epsfig{figure=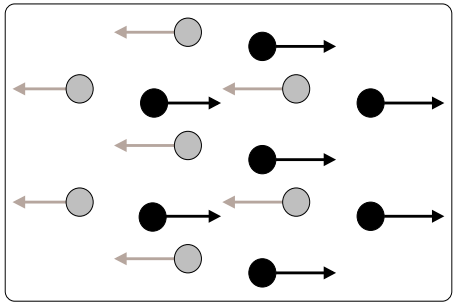,width=6cm,height=4cm}
\caption{Schematic diagram of the model} \label{model}
\end{figure}

We consider two sets A and B of Brownian particles in two spatial dimensions,
driven in the +x and -x directions respectively by a constant driving force
with magnitude $F$ as shown in Fig.~\ref{model}. Pairwise interactions between
particles are described by potentials $V_{AA}(r)$, $V_{BB}(r)$ and $V_{AB}(r)$.  
We choose a rectangular box of dimensions $L = (\sqrt{3}/2)\times 20 \ell$ 
and $W = 20\ell $, where $\ell= (2\sqrt{3}n_0)^{-1/2}$, $n_0$ being
the mean number density of either species. All quantities we use
are in nondimensional form. Lengths are non-dimensionalized by
$\ell$ and time by $\tau \equiv \ell^2/D$,$D$ being the Brownian
diffusivity. Energy is scaled by $k_BT$ and force by $k_BT/\ell$,
where $T$ is the temperature and $k_B$ the Boltzmann constant. 
If the two close packed planes, with species A and B in distinct planes, are
slid past each other, it is highly plausible that $V_{AB}$ will be weaker than
$V_{AA}$ and $V_{BB}$ (since the effect of the {\em interlayer} interaction on the
two-dimensional in-plane motion of the particles, which is what $V_{AB}$
encodes, is substantially smaller than that due to the {\em intralayer} interaction)
and that the strength of $V_{AB}$ relative to $V_{AA}$ and $V_{BB}$ can be varied by increasing or decreasing the normal confining 
pressure. We therefore choose, for this work dimensionless pair potentials of 
the screened Coulomb form
\begin{equation}
V_{AA}(r) = V_{BB}(r) = \epsilon^{-1}V_{AB}(r) = (U_0/r)
\exp(-\kappa r),
\end{equation}
where $U_0$ ($=1.75 \times 10^4$) and the screening parameter
($\kappa \ell=0.5$ ) are so chosen that each species in the
absence of the other and any external driving force settles down
in a triangular lattice configuration (Fig.~\ref{model}). The
particle positions then evolve according to the overdamped Langevin equations
\begin{eqnarray}
\label{e.langnondim} {\R_{\mu}}^i(t + \delta t) = {\R_{\mu}}^i(t)
+  \delta t [{\bf F_{\mu}}^i + {\bf f_{\mu}}^i({\R_{\mu}}(t)) +
{\bf h_{\mu}}^i(t)],
\end{eqnarray}
where ${R_{\mu}}^i=(x_{i\mu},y_{i\mu})$ is the coordinate of the
ith particle of the species $\mu$ ($=A$ or $B$), and $\pm F \hat{x}$ 
(+ for A, - for B) is the external driving force on the $i$th particle
of type $\mu$. 
\begin{equation}
{\bf f_{\mu}}^i(\R_{\mu}^i) = - \sum_{j \neq i} \nabla V_{\mu^i
\nu^j} (\R^i - \R^j)
\end{equation}
are the interparticle forces, and ${h_{\mu}}^{(i)}$  are Gaussian white noise 
sources with zero mean obeying the fluctuation dissipation relation which in
nondimensional form reads
\begin{equation}
\langle {\bf h_{\mu}}^i(0) {\bf h_{\nu}}^j(t) \rangle = 2 \Itens
\delta_{\mu \nu}\delta^{ij} \delta(t).
\end{equation}
It is trivial to generalize to the case where species A and B differ
but we have chosen them to be same in this model. We study the time evolution 
of this system as a function of the drive keeping $\epsilon$ constant for 
several values of $\epsilon$. The dimensionless time step used in our
simulations is $\delta t =6.4 \times 10^{-6}$.

\subsection{Simulation Results}

The results reported in our study are generally for  $10^6-10^7$
time steps after the initial transients $\sim$ $10^4$ steps are
discarded. Over this time, the A and B lattices sweep through each
other a few to several hundred times depending upon the magnitude
of the drive. In order to drift under the action of the driving
force $F$, the particles have to overcome a
barrier of the order of $V_{AB} (\ell)$  arising from 
interaction with the nearest neighbors of the opposite species.
Thus, although $F$ is itself dimensionless, it is appropriate to
state the results in terms of the physically relevant
dimensionless combination $F_d \equiv F \ell /V_{AB}(\ell)$.
However, for the phase diagram in the $\epsilon$-$F$ variables, we
have used the dimensionless combination ${F_d}^* \equiv F \ell \epsilon
/ V_{AB}(\ell)$, as $V_{AB}$ already incorporates a factor
of $\epsilon$ in its definition. The structure and dynamics of the
system has been monitored through snapshots of configurations,
drift velocities $v_d$, particle-averaged local velocity variances
$\langle (\delta v)^2 \rangle$, pair correlation functions $g_{\mu
\nu}({\bf r})$ as functions of separation ${\bf r}$, and time dependent,
but equal time structure factors $S_{\mu \nu}({\bf k},t)$ as functions of
wavevector ${\bf k}$ ($\mu$ and $\nu$ range over A, B). Each point
in the $g_{\mu \nu}({\bf r}), ({\bf r}=x,y)$ is an average over
100 data points, recorded at times separated by $50 \delta t$. In
the absence of the driving force ( i.e., at $F_d=0)$, the system
is an imperfectly ordered crystal. The application of a small
nonzero $F_d$, well below the apparent threshold for perceptible 
macroscopic relative
motion of the two lattices (of A and B particles respectively),
facilitates particles that are initially in unfavorable positions
to re-organize and move to favorable locations leading to a small
movement in these regions. After these transient motions the
system settles down into a macroscopically ordered structure with
both components showing  perfect long-range crystalline
order sustained over distances of the order of the system size.
There is no further relative drift of A and B except perhaps a
tiny activated creep which we cannot resolve. Thereafter, keeping 
interaction strengths
and temperature fixed, the driving force $F$ displays three
threshold values $F_i$, $i = 1, 2,3$ corresponding to the lower
bounds of three states -a lower sliding crystalline state, a
melt-freeze state and an upper sliding crystalline state. The
characteristic features of each of these three states are mentioned
below.

\begin{figure}
\epsfig{figure=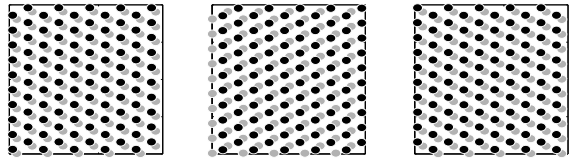,width=8cm,height=2.5cm}
\caption{Simulation images of macroscopically ordered lattices
drifting through each other for $\epsilon=0.05$, $F_d^{*}=0.0438$.}
\label{slidingorder}
\end{figure}

1. When $F_d$ crosses the first threshold $F_1$, the A and B
components acquire a measurable, macroscopic relative drift
velocity $v_d$. The drift velocity shows a smooth change at this
threshold value with the velocity fluctuations \cite{v1} showing a
pronounced enhancement \cite{msg} characteristic of depinning.
(See Fig. 3 and 4 of Ref.\cite{msg}.) This is likely to be a
strong crossover rather than a true transition. Each particle
faces a finite barrier to motion, so that at any nonzero temperature, 
particles can cross the barrier individually in an incoherent manner for 
arbitrarily small $F_1$. The barrier for creep velocity should thus be
finite even in the limit of infinite system size. In the region
$F_1 < F < F_2$, both A and B components are well-ordered,
drifting crystals (Fig.~\ref{slidingorder}). The two components slide
smoothly past each other in lanes of width equal to the
inter-particle distance, with negligible distortion or disorder. We
comment later on the connection to the lane formation work of \cite{lowen}.

\begin{figure}
\epsfig{figure=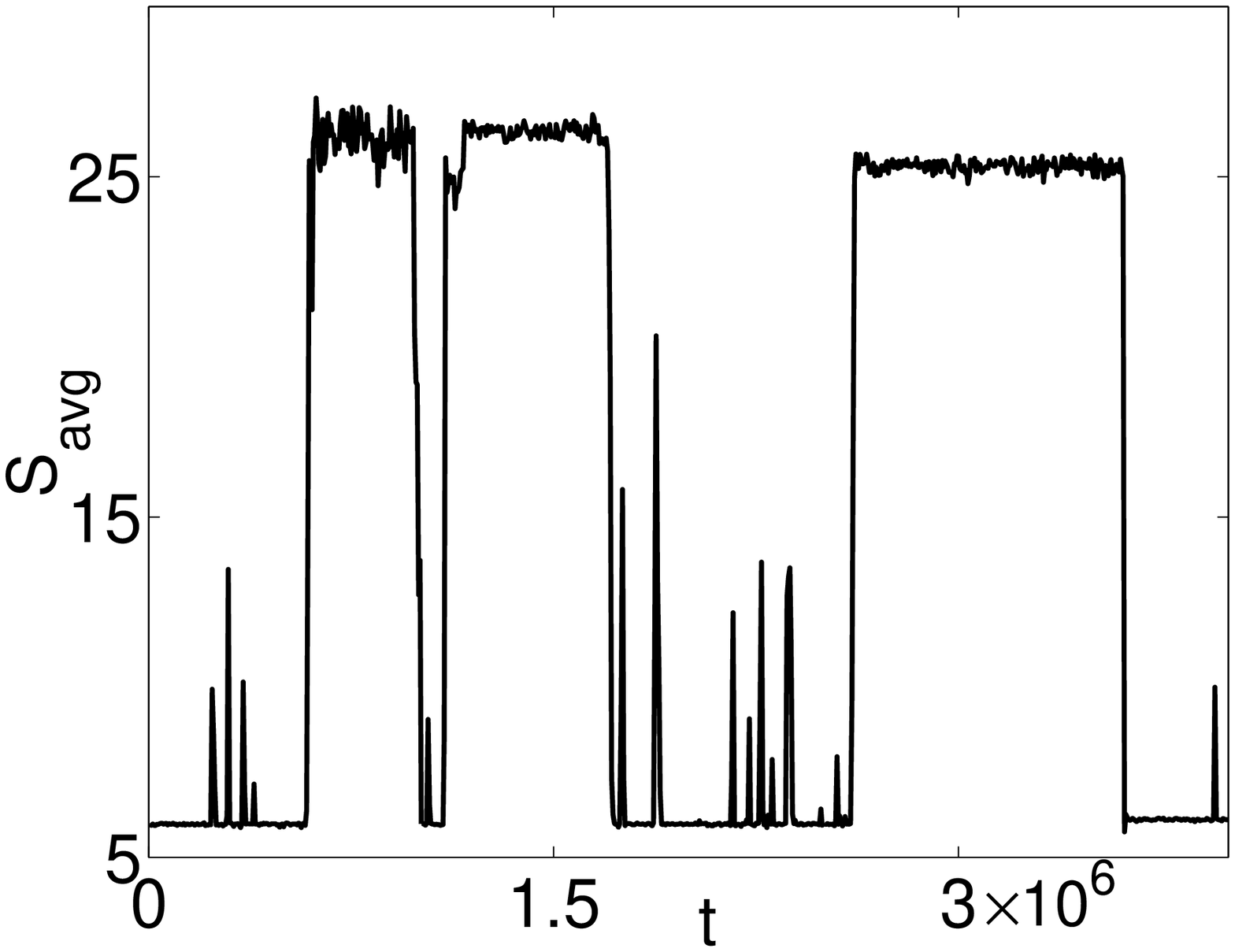,width=8cm,height=5cm}
\caption{The structure factor height (averaged over 1st ring of
maxima) as a function of time in the melt-freeze cycle state for
$\epsilon=0.05$, ${F_d}^{*}=0.8167 $.} \label{skmedium}
\end{figure}    

\begin{figure}
\epsfig{figure=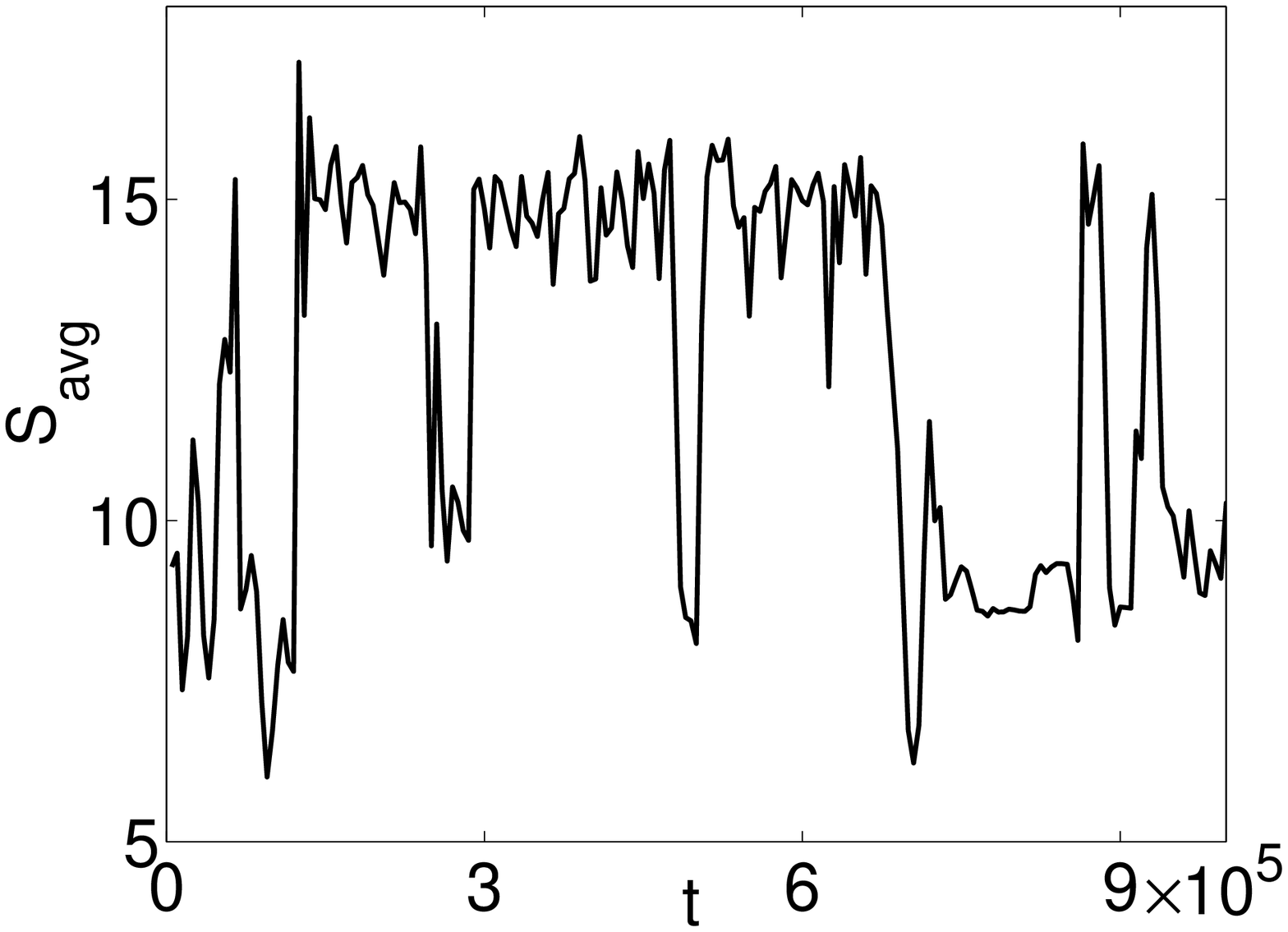,width=8cm,height=5cm}
\caption{The structure factor height (averaged over 1st ring of
maxima) as a function of time in the melt-freeze cycle state for
$\epsilon=0.02$, ${F_d}^{*}=0.8167$.} \label{sklow}
\end{figure}

\begin{figure}
\epsfig{figure=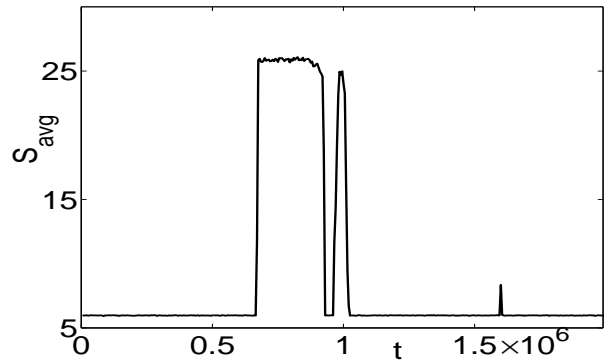,width=8cm,height=5cm}
\caption{The structure factor height (averaged over 1st ring of
maxima) as a function of time in the melt-freeze cycle state for
$\epsilon=0.06$, $F_d^{*}=1.4350$.} \label{skhigh}
\end{figure}

\begin{figure}
\epsfig{figure=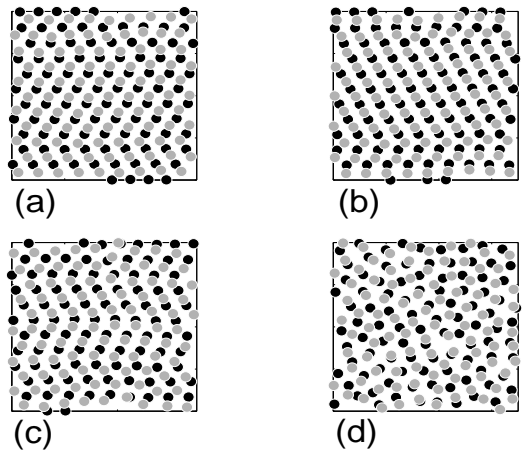,width=7cm,height=6cm}
\caption{Particle configuration snapshots at the onset of disorder
in the melting-freezing regime for $\epsilon=0.05$, $F_d^{*}=0.8167$.} \label{stripemelt}
\end{figure}

\begin{figure}
\epsfig{figure=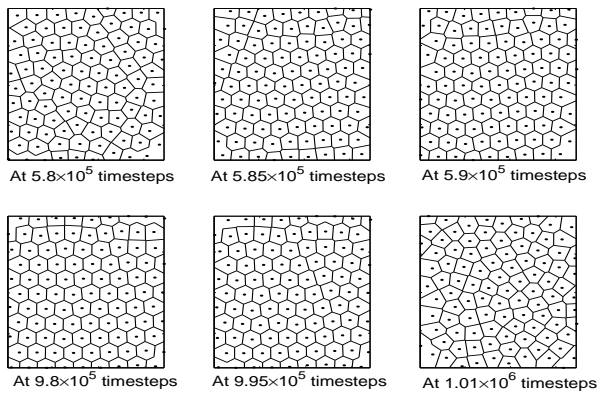,width=8cm,height=5.2cm}
\caption{Voronoi construction from particle configurations showing
the onset of order and of disorder for species $A$ at
$\epsilon=0.05$, ${F_d}^{*}=0.8167$.} \label{onsetvoronoi}
\end{figure}

\begin{figure}
\epsfig{figure=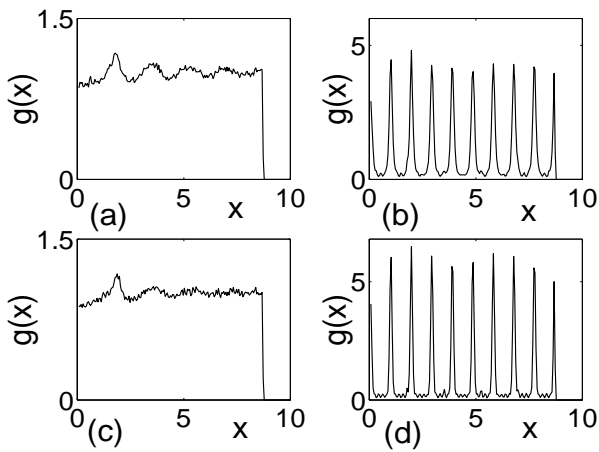,width=8cm,height=6cm}
\caption{The distribution function along $x$ direction for species
$A$ for $\epsilon=0.05$, ${F_d}^{*}=0.8167$.} \label{gxaa}
\end{figure}

\begin{figure}
\epsfig{figure=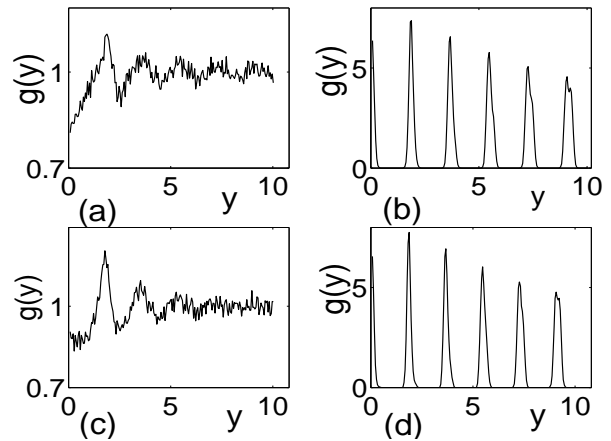,width=8cm,height=6cm}
\caption{The distribution function along $y$ direction for species
$A$ for $\epsilon=0.05$, ${F_d}^{*}=0.8167$.} \label{gyaa}
\end{figure}

2. In a window of driving forces $F_2 < F < F_3$, we observe 
intriguing stochastic alternations of the system between an
ordered and a disordered state  which are the melt-freeze cycles 
(Figs.~\ref{sklow},~\ref{skmedium},~\ref{skhigh}).
We have used  snapshots of configurations Figs.~\ref{stripemelt} and ~\ref{onsetvoronoi}, 
the pair correlation function $g_{\mu \nu}(\bf r)$,${\bf r} = x, y$ 
(Figs.~\ref{gxaa} and ~\ref{gyaa}) and the
equal-time but time-dependent structure factor $S({\bf k}, t)$ to 
characterize these dynamical states of order and disorder. As the first 
two methods are representative of the instantaneous state of the system, for
monitoring these cycles continuously we use the peak height of the
(short-time averaged) static structure factor $S({\bf k},t)$.
Indeed, the essential features of these cycles, namely the
persistence duration, fluctuations in the extent of order etc, are
best captured through the time dependence of $S({\bf k},t)$ as we
shall see. A typical such plot is shown in Fig.~\ref{skmedium} for an
optimum value of $\epsilon =0.05$. It is clear that $S({\bf k},t)$
alternates between long stretches of crystal-like ( corresponding
to large values of $S({\bf k},t)$) and {\it comparably} long stretches
of liquid-like values ( small values of $S({\bf k},t)$ ) as the
simulation progresses. This is strikingly different from stick-slip alternations, in which the melted (slip) state is considerably shorter than the crystalline (stick) state \cite{homola,thompson}. We discuss this comparison later in section IV. 
For smaller values of $\epsilon$, even as
the persistence of the crystalline state is enhanced, the extent
of ordering itself is less pronounced as indicated by
the lower values of $S$ ( compared to that for the crystalline
phase corresponding to $\epsilon = 0.05$). Concomitantly the
liquid-like state also displays significant short range order. 
These features are clear from Fig.~\ref{sklow} where $S ({\bf k},t)$ is 
shown for $\epsilon = 0.02$. For higher values of $\epsilon$, the
liquid-like state is more favored (as can be seen from the short
stretches of crystalline order) for $\epsilon = 0.06$ shown in
Fig.~\ref{skhigh}.

As is clear from Figs.~\ref{skmedium},~\ref{sklow},~\ref{skhigh}, 
the time scales over which the crystalline order
or liquid-like disorder sets in is considerably shorter than the
persistence time of each of these phases. The process of ordering
and disordering is better monitored through a sequence of
snapshots of configurations. A typical set of snapshots are shown
in Fig.~\ref{stripemelt}. The corresponding Voronoi construction for
$\epsilon = 0.05$ is shown in Fig.~\ref{onsetvoronoi}. ( See also Figure
5 in Ref.\cite{msg}.)  The extent of order in the $x$ and $y$
directions is studied using the correlation function. The nature
of $g(x)$ and $g(y)$ is shown in Fig.~\ref{gxaa} and ~\ref{gyaa}
at four different times. Note that the extent of order in the
liquid-like state in the direction of the drive x is significantly less than 
that in the $y$ direction. Further the primary ordering wavevectors are
along $\hat{x}$ and $60^{\circ}$ to $\hat{x}$. Thus nearest neighbor distance
in the y direction is $\sqrt{3}$ times larger. A partial understanding of why
order starts to set in again after melting has occurred is to note that relative
motion disrupts primarily those structures with ordering wavevector along the 
drift direction $\hat{x}$, leaving some residual order along $\hat{y}$ as
seen in Figs.~\ref{gxaa} and ~\ref{gyaa}. So each
species still provides a weak periodic potential along $y$ for the
other species. This can induce order along $x$ as well resulting
in a 2d ordered state by a mechanism similar to the ``laser induced
freezing'' of a 2d suspension of strongly interacting colloidal
particles subject to a 1d periodic modulation 
\cite{chowdhury,barrat,cdas,leo}. This state persists for a long
time, before disorder once again sets in. And it is in this
driving force regime that one sees the melting-freezing cycles.
Finally, we find that the two species do not necessarily order or
disorder simultaneously. Though we found no clear trend in this
regard, mostly, one species begins to order while the other is
disordered. Also, for both species, the `Bragg peak heights' rise
fast and decline slowly, i.e., ordering takes place much faster
than the progression of disorder.


\begin{figure}
\epsfig{figure=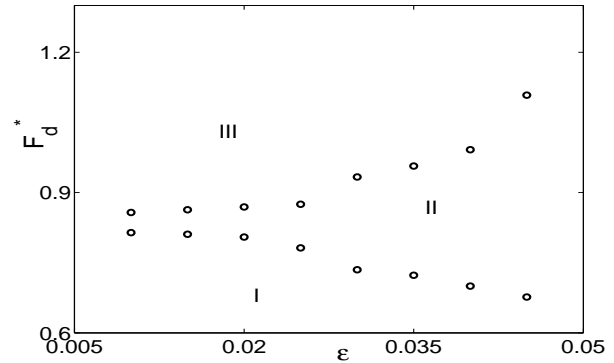,width=8cm,height=5cm}
\caption{The phase diagram of the system (100 particles of each species)
in the $\epsilon$-${F_d}^*$ plane, where ${F_d}^*=F\ell\epsilon/V_{AB}$.
The system undergoes the stochastic 'melt-freeze' alternations in
the region II, and is a macroscopically ordered crystal in the the
upper and lower regions I and III respectively.} \label{phase1}
\end{figure}

The structure factor can be used to obtain the dynamical phase
diagram in the $F_d^{*} -\epsilon$ plane. This is shown in 
Fig.~\ref{phase1}. The points shown here actually represent the value
of $(F_d^{*}, \epsilon)$ for which the crystalline phase is
detected just before entering the bistable melt-freeze region (
region II). The regions I and III refer respectively to the lower
and upper sliding crystalline states. We emphasize here that the
transition to the melt-freeze regime occurs over a finite range of values 
of the parameters and it is not possible to pin down with precision
the point at which the melt-freeze alternations begin. The range of values of
the force, $F_3-F_2$, over which  we observe the melt-freeze cycles
increases with $\epsilon$ and hence with $V_{AB}$ as well 
(Fig.~\ref{phase1}). This agrees well with our observation that the
average potential barrier that a particle has to negotiate during
its motion in the steady sliding state increases with $\epsilon$.
Further, for large values of $\epsilon$ ($\epsilon \ge 0.05$), the
alternations persist over a very large window of driving forces.
For such $\epsilon$ values, we have not been able to detect the
upper threshold $F_3$ corresponding to the reentrant crystalline
state.

We shall now discuss the melt-freeze cycles in more detail and
explain  the observed features. There is a curious metastability
associated with the cycles: for the parameters mentioned above, a
disordered configuration fails to nucleate even over our longest
simulations if the initial state is chosen to be a {\em perfectly
ordered lattice}. Thus, both the melt-free cycles and ordered
sliding states display local dynamical stability. But if we
disturb this initial perfectly ordered lattice by moving a single
particle by, say, one lattice spacing, the melt-freeze cycles
resume. Also note that the orientation of the triangular lattice in the ordered state
of the cycles (Fig.~\ref{stripemelt}(a)) is changed by $30^{\circ}$
with respect to the one in the steady sliding state (Fig.~\ref{slidingorder}). 
This exchange of stabilities between the fcc-like (Fig.~\ref{slidingorder}) and "layer"
(Fig.~\ref{stripemelt}(a)) structures is known from experiments \cite{pusey} and simulations
\cite{nose}. At low relative velocity, particles in each layer have ample time to get out of 
the way of those in the other layer while retaining on average the fcc-like structure. 
As the speed is increased, the structures do not have sufficient time to relax and overlap is 
reduced by going to the "layer" structure of Fig.~\ref{stripemelt}(a).  However, we observe 
that even with such a shift in orientation, in the ordered part of the cycle, the smooth 
relative motion of the A and B lattices is disturbed now and then by kinks;
a row moving out of step with adjacent rows, as marked in 
Fig.~\ref{stripemelt}. This leads to the formation of kink-like undulations
transverse to the mean drift like a wave. At some point in time as
these undulations build up sufficiently, the system enters a
disordered state. This state persists for a long time, before
order sets in once again. Recall that at small $\epsilon$, in the
ordered part of the cycle, the magnitude of the structure factor
is small and correspondingly we find  an enhanced level of
fluctuations (Fig.~\ref{sklow}) compared to that at large
$\epsilon$ (Fig.~\ref{skmedium} ). This is because the ordered states
at small $\epsilon$ can support a larger number of defects without
making a transition to the disordered state, whereas as $\epsilon$
is increased such states can not be sustained for long time. In
fact, with increasing $\epsilon$, the probability of the system
being in the ordered state crucially depends on the defect
density, i.e., the ordered state is long-lived only when the
number of defects is small. This can be understood by considering
the potential landscape  seen by each species. In the ordered
state, particles of each species are nested in the three-fold
minima formed by its nearest neighbors of the other species. For
small $\epsilon$, both in the lower and upper part of the sliding
states, the potential depth is shallow and creation of defects
in shallow potentials does not cost much energy. For the same
reason, the number of such defects that the state can sustain can
also be large which in turn implies that the extent of order in
the crystalline part of the cycles is not significant. A shallow
potential also allows for the ease of annealing of the defects as
this can be accomplished by removing one particle from a 7-fold 
co-ordinated or adding one particle to a 5-fold coordinated site 
(Fig.~\ref{onsetvoronoi}). 
Thus, low $\epsilon$ situation allows for the ease of
creation and annealing out of these defects as time progresses.
This dynamical balance between the creations and annihilation of
the defects  can be sustained for long stretches of time but with
smaller extent of order with concomitantly large fluctuations.
This is precisely what is seen in Fig.~\ref{sklow}. (Note that
this picture is also consistent with the observed feature that the
liquid-like state at low $\epsilon$ has a fairly high level of
order compared to that at high $\epsilon$.) In contrast, with
increase in $\epsilon$, the well depth increases significantly
which implies that the crystalline order would be high as can be
seen in Fig.~\ref{skmedium}. Moreover, the formation of defects is
less favored, but, once formed, it cannot be easily annealed.
Further each defect gives rise to large local restoring forces 
and the crystalline order will be terminated even
when their number is small.

3. For $F > F_3$, both A and B components are once again
well-ordered, drifting crystals. In fact, the re-appearance of a
smooth sliding state is akin to the the re-entrant ordered state
seen in \cite{mark,rlsrsmelt}.

In fact the sliding crystalline states that we observe at low and high drives
can be visualized as ordered states with lanes \cite{lowen} of single-particle 
width. Ref.\cite{lowen} studies a model very similar to ours, with $\epsilon=1$.
In that case, the equilibrium state is a crystal, randomly occupied by each 
species and the driven state shows the interesting phenomenon of lane 
formation. Because $\epsilon$ is large, the system tries to minimize the extent of AB interface, hence one gets a few broad lanes with many columns of 
particles of the same species. Since $\epsilon$ is very small for us, we get 
many lanes of unit width.  

The results stated above are for 100 particles of each species. 
We have carried out a systematic study of this phenomena for
144, 169 and 256 particles of each type for $\epsilon=0.02$. 
We find the same qualitative behavior as that for 100 particles 
including the range of $F_d$ values for the three phases. 
However, for smaller systems (with $64$ particles of each
species), we have not observed any significant decay in order for
$\epsilon < 0.1$ (possibly due to the fact that the correlation
length is of the order of half the system size in this case).

We now discuss some natural timescales which will be useful in our 
final explanation of this phenomenon. When the
two arrays of particles are driven through each other, there is a
competition between two timescales, $\tau_1 $ the time to traverse
one lattice spacing and $\tau_2 $ the timescale of relaxation of a
particle in the local potential well provided by its neighbors.
For $\tau_1 >> \tau_2$, each species has ample time to relax to local
equilibrium, while for  $\tau_1 << \tau_2$, each species averages out 
the undulating landscape. For both $\tau_1/\tau_2$ $>> 1$ 
and $<<1$, we expect and find smooth, orderly sliding. 
 We therefore expect maximal effects of interspecies interaction where $\tau_1$
$\approx$ $\tau_2$ (for example, for $\epsilon=0.05$, $\tau_1 \simeq 0.00022$ and 
$\tau_2 \simeq 0.00019$ in the 'melt-freeze' regime \cite{msg}) 
, which is what we find. This suggests that a detailed
explanation lies in mechanisms involving competing timescales to which we
now turn.

The stochastic alternations of the system between the crystalline 
and liquid-like states is strongly reminiscent of the phenomenon
of stochastic resonance (SR) \cite{benzi,nicolis,stochres,wisen}. 
To see  the similarities and the
differences, consider a prototypical example of stochastic resonance
of a Brownian particle in a bistable potential subjected to a weak
periodic forcing term. When half the periodicity of the driving
force is comparable to the mean first passage time associated with
the barrier crossing, the state of the system switches between the
two minima of the potential in a surprisingly regular way. Indeed,
the time series of the position of the Brownian particle entrained
in the two minima is  very similar to $S({\bf k},t)$ of the
particle model. What this potential ( or an `effective
free-energy' in this case) is and what modulations are, is not
clear in the present model and needs further investigation. One
can anticipate that the role of the bistable potential in SR is
played by the `effective free energy' as the system is a many
particle system switching between the crystalline and liquid-like
states. However, identifying a periodic forcing in the present
context is more difficult. Thus, it would be useful to construct a
reduced model which displays the dominant features of the particle
model. One important characteristic feature of SR is that the
signal to noise ratio exhibits a maximum at an optimum value of
the  noise intensity. This aspect cannot be easily checked in the
particle model as it involves generating very long time series
(which would involve prohibitively large scale computing ).
However, we recall that altering $\epsilon$ in the particle model
has an effect that controls the ratio of the residence times of
the system in the ordered and disordered states. This is similar
to altering the height of the bistable potential which in turn
controls the residence time in the example considered. This
identification further supports our view that the melt-freeze 
cycles are in fact stochastic resonance. We shall make
this more concrete by introducing a reduced model which captures
most  features of the particle model.

\section{THE REDUCED MODEL}
The effects of external nonequilibrium driving conditions in an
underlying first order phase transition have often been studied
successfully by modifying, say, the time dependent Ginzburg-Landau
equation for the dynamics of the order parameter \cite{onuki}.
The results of the previous section were obtained from
direct simulation of particle motion. In this section,
we propose an understanding of these results through
dynamical equations for the appropriate order
parameter fields evolving under the combined
effect of shear and a coarse-grained free energy. The
nature of the free energy functional is usually determined based
on the knowledge of the allowed states of order/disorder and a few
general symmetry considerations. Here, we follow the model proposed 
earlier \cite{rlsrsmelt} for studying sheared colloidal crystals 
\cite{coshear,mark}. Recall that in our simulations, at low drives, we
find a smooth sliding crystal wherein the A lattice slides  past
that of $B$ in a coherent fashion, and at intermediate drive
values, we observe the melt-freeze cycles. To mimic this, we
choose  an order parameter denoted by $\rho$ (the Bragg peak  
intensity), which takes on a finite value corresponding to
the crystalline order and zero value corresponding to the
liquid-like order. A simple form of the free energy which ensures
the crystalline ($\rho \neq 0$) and melt phases ($\rho= 0$) is the
Landau polynomial  for a first order transition
\begin{equation}
V(\rho)=\frac{a_1 \rho^2}{2} - \frac{b_1 \rho^3}{3} + \frac{c_1
\rho^4}{4}.
\end{equation}
The distortions produced due to the drive in the particle model
can be represented by another variable representing the strain
(in our model, the relative phase of the density wave in the two sliding
layers) denoted by $\theta$. In the crystalline state, as distortions are
small and homogeneous at low drives, we take $\theta$ to be zero
for this state. As homogeneous distortions would mean that the
successive minima of the crystal are equivalent, we consider
$\theta$ to be a periodic variable with $\theta = 0$ to be
equivalent to $\theta = 1$. Further recall that our simulations
show that at intermediate drives, deformation becomes
inhomogeneous forcing the crystalline state to melt ( although in
a dynamical way).  Again, a simple form of the free energy in
$\theta$ should incorporate elasticity at small $\theta$ and
yielding at large $\theta$, ie., at small strains, $V(\theta)$ is
assumed to be quadratic in $\theta$ and a softening term at larger
strains ( the cubic term). The coefficient of $V(\theta)$ must vanish
with $\rho$, say as $\rho^2$, as the free energy cost of deformations 
must reduce to zero when the system is in the liquid state (i.e., for $\rho=0$).
The simplest general form of the free energy $F(\rho,\theta)$ of a distorted 
solid respecting the above
conditions is \cite{rlsrsmelt} of the following form
\begin{equation}
F(\rho,\theta)=V(\rho) + \frac{1}{2} {\alpha}
\rho^{2} V(\theta),
\end{equation}
where $V(\theta)$ has a similar form as $V(\rho)$
\begin{equation}
 V(\theta)=\frac{a_2 \theta^2}{2} - \frac{b_2 \theta^3}{3} +
\frac{c_2 \theta^4}{4}.
\end{equation}
 The Langevin-TDGL equations for $\rho$ and $\theta$ that describe the 
dynamics of this system are
\begin{eqnarray}
\dot{\rho} =
-\frac{1}{\Gamma_{\rho}}\frac{\partial{F(\rho,\theta})}{\partial{\rho}}
+ \eta_{\rho},
\label{glreqn}
\\
\dot{\theta} =
-\frac{1}{\Gamma_{\theta}}\frac{\partial{F(\rho,\theta})}
{\partial{\theta}} + \Omega + \eta_{\theta}.
\label{gleqn}
\end{eqnarray}
The idea is that in the absence of any restoring forces for $\theta$, 
$\dot{\theta}$ would be equal to $\Omega$. In general, then, $\Omega$ 
represents the effects of relative sliding of the layers, at a rate 
determined by competition between $\Omega$ and $\frac{\partial F}{\partial{\theta}}$. 
If $\Omega$ is too small, $\theta$ will get stuck at a finite value in the 
absence of noise. ${\Gamma_{q}}^{-1}$'s $ (q=\rho,\theta)$ are the kinetic 
coefficients, $\eta_q$'s
represent Gaussian delta-correlated noise components whose
variances are related to $\Gamma_q$ and temperature through the
fluctuation-dissipation relation. We ignore possible additional 
non-equilibrium noise sources.

The equations for our system are those for an overdamped
particle in an nonsymmetric double well potential
$F(\rho,\theta)$, driven along the angular coordinate \cite{f1}.
For zero strain ($\theta=0$), the system relaxes in either of the
two minima corresponding to the liquid minimum
\begin{equation}
\rho_{l}=0,\theta = 0,
\end{equation}
or the crystalline minimum
\begin{equation}
\rho_{c}={\frac{b_1}{2c_1}}\left [1 + \sqrt{1-\frac{4a_1c_1}{{b_1}^2}}\right ],
\theta =0,
\end{equation}
depending on which is the locally stable state.

We choose the parameter values ($a_1,\, b_1$ and $c_1$) such that
the crystalline minimum $\rho_{c}$ is the more favorable state at
zero drive and the potential barrier between the two minima,
$V(\rho_{l}) - V(\rho_{c})$, is appropriate. In the presence of
noise ( whose strength can be appropriately chosen) the system
equilibrates with the respective populations determined by  noise
strength and the relative well depths. If $\theta$ and $\rho$ are 
macroscopic (infinite system-size) averages, there should be no noise
in the equations. In practice, presumably, shear melting and the cycles
take place over a finite correlated domain (whose size we do not know). 
We are therefore justified in using a noisy evolution equation. 

When the drive $\Omega$ is switched on, this scenario is altered
and the populations of each of these wells now evolve in time
depending upon the competing time scales of relaxation and applied
shear rate. In this case, it is better to consider the fixed
points of the noise free case of Eq. \ref{glreqn} and \ref{gleqn}.
The two attractive fixed points to which the system relaxes  can now
be identified with crystalline and liquid-like order. The
repulsive fixed point determines a saddle type of maximum. These
will depend on the shear rate $\Omega$. The barrier height between
the stable fixed points and unstable fixed point determine the
barriers that the system has to surmount. These now depend on
$\Omega$. We find that the value of the free energy at the liquid
like minima and the saddle do not change significantly, only the
crystalline (distorted) minimum changes as a function of  $\Omega$
given by
\begin{equation}
\rho_{c}={\frac{b_1}{2c_1}}\left [{1 + \sqrt{1-\frac{4(a_1+\alpha
V(\theta_{min}) )c_1}{{b_1}^2}}}\right ].
\end{equation}
Note that Eq. \ref{gleqn} determines a critical value of $\Omega
=\Omega_c = 0.5 \alpha \rho^2 \frac{\partial V}{\partial \theta}
\vert_{max}$. For values of $\Omega < \Omega_c$, the  barrier is
high. In such a situation, in the presence of noise, the
transitions would be rare. But, on increasing $\Omega$ beyond the
critical value the ``free energy'' of crystalline state becomes
comparable with that of the liquid minimum. Under these
conditions, noise assisted transitions to the liquid state occur.
More importantly, in this regime, as $\theta$ is itself changing
as a function of time, the minima in $(\rho,\theta)$ is slowly
modulated under action of the drive $\Omega$ and the relative
stability changes as a function of time. When the time scale
imposed by $\Omega$ is small enough to allow the system to make
inter-well transitions and when the time scale of the induced
periodicity is approximately equal to the Kramers escape time 
\cite{kramers,krmp} under the influence of noise, one expects transitions 
between the crystalline and liquid like states in a range of values of
$\Omega$ beyond $\Omega_c$. As a result the system can undergo
stochastic transitions between the two metastable states which in
turn can lead to comparable lifetimes.\\

As this is a driven system, a reasonable criterion for studying
the occupancy of the system  is to calculate the marginal
probability distribution function $P(\rho)= \int P(\rho,\theta)
d\theta$ (i.e., the probability of the order parameter having the
value $\rho$, independent of the value of the strain field
$\theta$). In the following section we show that the external
drive causes the system to sample both minima or stay mainly in 
one of the minima depending on the value of the drive $\Omega$, noise
strength, coupling constant $\alpha$ starting from an initial
crystalline
state.\\

\subsection{Results of the Reduced Model}

We study the time series and the probability distribution of the
system by discretizing the Langevin equations in time. The
integration scheme used is the fourth order Runge-Kutta with a
fixed time step of 0.001. After discarding  transients ($\sim
10^5$ time steps) the time evolution of the system for the next
 $8 \times 10^9$ time steps is monitored. In figures \ref{diffdrivetime}
and \ref{diffcoupltime} we have shown a timeseries stretch from $6 \times 10^9$ 
to $6.5 \times 10^9$ timesteps.  Noise corresponding to
$\rho$ and $\theta$ are drawn from  Gaussian white noise
distributions with zero mean and unit variance. For studying the
time evolution of the system, we have chosen a noise strength $D=7
\times 10^{-4}$ for both $\eta_{\rho}$ and $\eta_{\theta}$. ( We
shall study the influence of noise on the signal to noise ratio
later.) For the numerical work reported here, the parameters
values in the free energy used are $a_1 = 0.85$, $b_1 = 5.8$,
$c_1=8.0$, and for the strain field are $a_2 = 1.3644$, $b_2=
8.7105$ and $c_2=13.6740$. We study the time evolution of the
system as a function of the drive $\Omega$ (for a fixed coupling
parameter $\alpha$) and as a function of $\alpha$ for a fixed
drive. We observe the following behavior:

a. {\underline{As a function of the drive $\Omega$ :}

Keeping $\alpha$ fixed at an optimum value ($\alpha = 0.17$), the
drive has two thresholds:

i. For $\Omega < \Omega_1$  and $\Omega > \Omega_2$, we find the
system always resides in the crystalline minimum ($\rho \ne 0$).
The time series of $\rho(t)$ is shown in Fig.~\ref{diffdrivetime}a,c for a
typical set of values $\Omega =0.03$ and $\Omega = 0.3$
respectively. The corresponding probability distribution is peaked
around the crystalline minimum, as can be seen from
Fig.~\ref{diffdriveprob}a,c.
\begin{figure}
\epsfig{figure=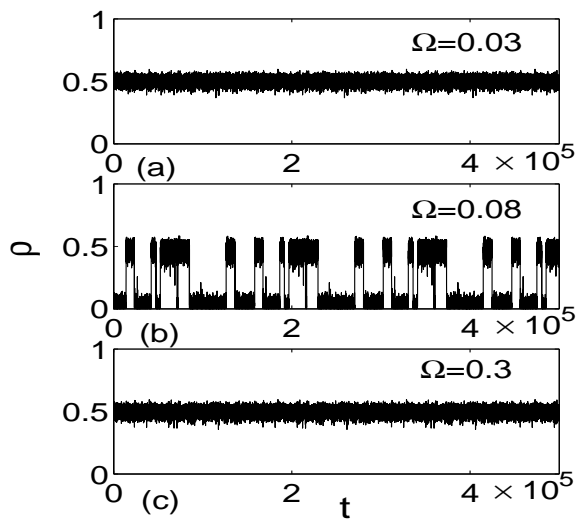,width=8cm,height=7cm}
\caption{The time series of the order parameter in the low (top
panel), high (bottom panel) and intermediate (middle panel)
driving force regimes. We have chosen a noise strength $D=7\times
10^{-4}$ and coupling parameter $\alpha=0.17$.} \label{diffdrivetime}
\end{figure}

\begin{figure}
\epsfig{figure=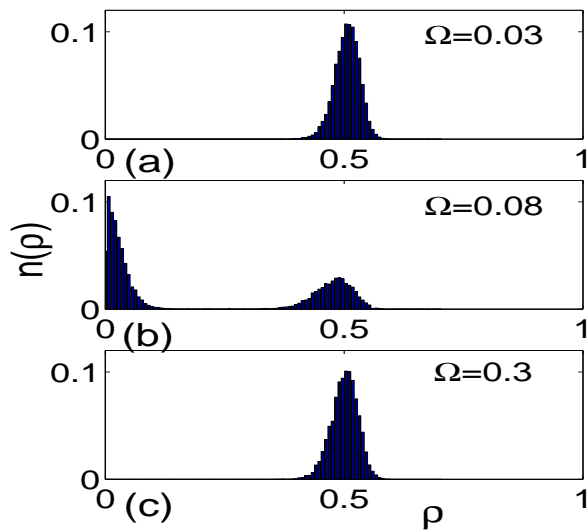,width=8cm,height=7cm}
\caption{The marginal probability distribution of the order
parameter in the low (top panel), high (bottom panel) and
intermediate (middle panel) driving force regimes for the same
parameter values as in Fig.11.} \label{diffdriveprob}
\end{figure}

ii. In an intermediate window of driving forces $\Omega_1 < \Omega
< \Omega_2$, the system stochastically alternates between the
liquid ($\rho=0$) and the crystalline minimum. The time series
corresponding to a typical value of $\Omega = 0.08$ is shown in
Fig.~\ref{diffdrivetime}b. The probability distribution has two peaks
corresponding to these two minima. The time series of the order
parameter (Fig.~\ref{diffdrivetime}b) also shows that the persistence time
of these two states are comparable (for optimum values of
$\alpha$) very much like the time series of a system undergoing
stochastic resonance. These results  are similar to the results of
the particle model where $F^*_d$ was varied for a fixed value of
$\epsilon$.

b. {\underline{As a function of the coupling $\alpha$ :}

Here, we have kept the drive $\Omega$ at intermediate values.  We
find that the crystalline minimum is favored over the liquid one
for low values of $\alpha$, whereas for high values the liquid
minimum dominates. At intermediate values of $\alpha$, the system
spends comparable durations in each state. As we increase $\alpha$
from small values, one finds that the system tends to spend
increasingly more time in the liquid state and, eventually, at large
values of $\alpha$ the liquid state is the preferred state. This
is shown in Fig.~\ref{diffcoupltime} for three typical values of
$\alpha$. (The numerical results are for  $\Omega = 0.08$ for
which we observe the most prominent stochastic switching of the
order parameter values between $\rho \ne 0$ and $\rho =0$.)
 This is also reflected in the probability distribution
which has a more pronounced peak at the crystalline minimum at
small $\alpha$ and at the liquid minimum at large $\alpha$. For
intermediate values of $\alpha$ we find a bimodal distribution.
Figure \ref{diffcouplprob} shows the probability distributions for the
parameter values of Fig.~\ref{diffcoupltime}.  This behavior is similar
to the results obtained in the particle model by varying
$\epsilon$ keeping $F^*_d$ fixed.

\begin{figure}
\epsfig{figure=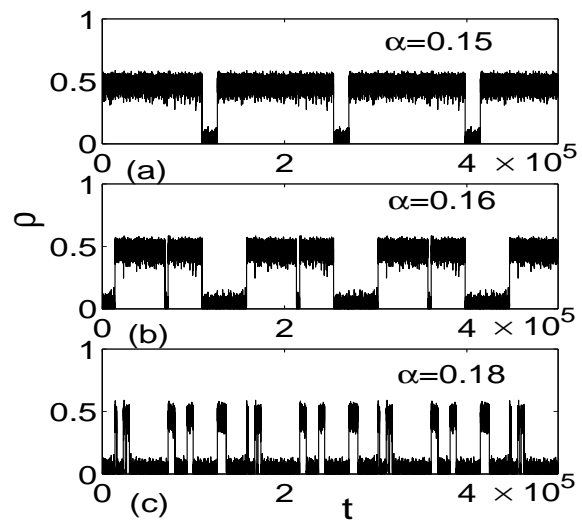,width=8cm,height=7cm}
\caption{The time series of the order parameter in the low (top
panel), high (bottom panel) and intermediate (middle panel) values
of the coupling parameter $\alpha$ for noise strength $D=7\times
10^{-4}$ and driving force $\Omega=0.08$.} \label{diffcoupltime}
\end{figure}
\begin{figure}
\epsfig{figure=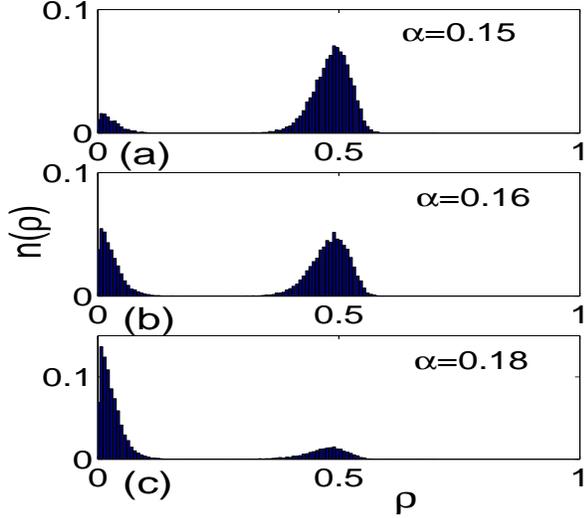,width=8cm,height=7cm}
\caption{The marginal probability distribution of the order parameter in
the low (top panel), high (bottom panel) and intermediate (middle
panel) $\alpha$, for the same parameter values as in Fig.13.}
\label{diffcouplprob}
\end{figure}
As mentioned, one dominant feature of SR is the  enhancement of
the signal to noise ratio (SNR) at an optimum  value of the noise
intensity. In order to check this, we have carried out long runs
of the order of $10^{10}$ time steps. The power spectral density
(PSD)  of the time series has been calculated for various values
of the noise intensity  $D = 6 \times 10^{-4}$ to $ 1.2 \times
10^{-3}$. It exhibits strong peaks at all integral values of the
fundamental (unlike the symmetric bistable potential where only
the odd harmonics are seen) due to the absence of any symmetry in
$F(\rho,\theta)$. A plot of this is shown in Fig.~\ref{power} for
a typical value of $D = 7.0 \times 10^{-4}$. 
The signal to noise ratio calculated
from the power spectrum using the first peak for various values of
$D$ is shown in Fig.~\ref{snr}. (Here we have used the conventional SNR
defined by $ SNR = 10 \, log [S_{signal}(\omega)/S_{noise}(\omega)$].)
As is clear from the figure, the maximum enhancement of the SNR is
found to be around $D \sim 7.0 \times 10^{-4}$.
\begin{figure}
\epsfig{figure=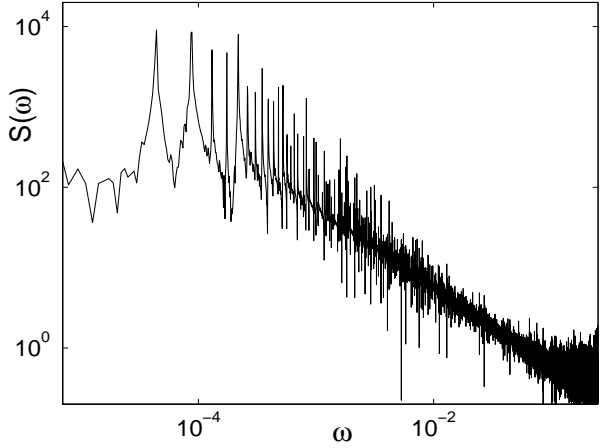,width=8cm,height=6cm}
\caption{The power spectrum of the time series of the order
parameter $\rho$ for $\alpha =0.17, \Omega =0.08$ and $ D = 7.0
\times 10^{-4}$.} \label{power}
\end{figure}
\begin{figure}
\epsfig{figure=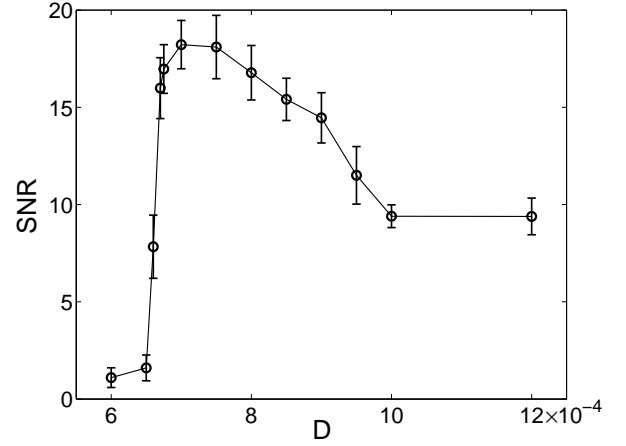,width=8cm,height=6cm}
\caption{The signal to noise ratio for the reduced model for
parameter values $\alpha =0.17, \Omega =0.08$.} \label{snr}
\end{figure}

As discussed above, we note that $\Omega$ and $\alpha$ of the
reduced model take the roles of the drive $F^*_d$ and the
interspecies interaction strength $\epsilon$  in the particle model
respectively. To make the parallel between these models more
concrete, we have constructed  the dynamical phase diagram in the
$\alpha-\Omega$ plane shown in Fig.~\ref{phase2}. ( In
constructing this diagram, we have taken the system to be a liquid
state if it spends less than $2\%$ of the time in the crystalline
minimum, and correspondingly for the crystal.) Region I refers to
the crystalline phase and region III, the re-entrant crystalline
phase. The melt-freeze cycles where the system alternates between 
crystal and liquid is shown as region II. In
region II, for low values of the coupling parameter $\alpha$, the
persistence of the ordered state is more than that for the
disordered state, and decreases as $\alpha$ is increased,
eventually giving way to the liquid-like region IV for moderate
values of $\Omega$. It is clear that this diagram is remarkably
similar to the phase diagram of the particle model shown in Fig.
~\ref{phase1}, except for the region IV which could not be detected
in the particle model within the longest runs carried out.
\begin{figure}
\epsfig{figure=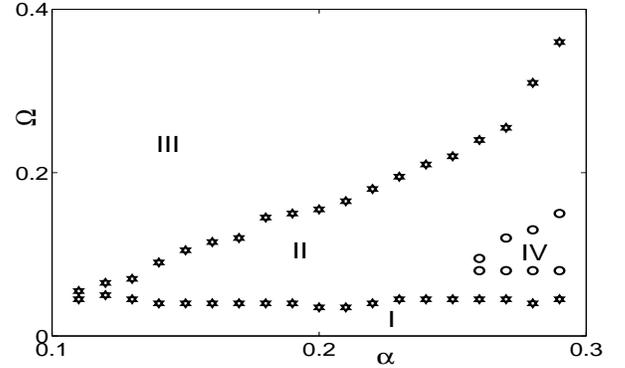,width=8cm,height=5cm}
\caption{The phase diagram of the reduced model in the
$\alpha$-$\Omega$ plane. The system is crystalline in region $I$
and $III$ represents the re-entrant solid. In region $II$ we
observe the `melt-freeze cycles'. For high values, liquid-like
region $IV$ is seen again.} \label{phase2}
\end{figure}

The physical picture of the reduced model is clear. As the initial
state at zero shear rate is a crystalline state, by continuity
arguments one should expect that at low shear rates the system
should find the crystalline minimum favorable. At intermediate
range of shear rates, the system develops another minimum at zero
value of the order parameter $\rho$ corresponding to the melt
state making the system bistable. When the shear rate is close to
( and larger than) the critical value $\Omega_c$,  under the
action of noise, the systems makes transitions from the
crystalline minimum to  the liquid minimum and vice versa.
However, since the strain $\theta$ evolves in time the system
experiences an additional periodic modulation. We note here, that
this periodicity is not equal to $1/\Omega$ as the the strain
variable moves on the $V(\theta)$ surface. Typical values of the
induced periodicity  estimated from deterministic version of Eqs.
~\ref{glreqn} and \ref{gleqn} for the optimum range of drive
values is of the order of $10^5$ time units. When the induced
periodicity is twice the Kramers escape rate \cite{kramers}, the system
alternates between the two minima. Further, we note that as the
two wells are not symmetric, in the presence of noise, the mean
first passage times associated with the two wells will be
different. Indeed, we find that as a function of $\Omega$, there
is a range of $\Omega$ values ($0.065 < \Omega < 0.12$) where the
barrier between the crystalline minimum and maximum of the   
free energy $F_{saddle}-F_{crystal} = \Delta F_{c}$, is much smaller
than that between the liquid like minimum and the maximum,
$F_{saddle}-F_{liquid} = \Delta F_{l}$. A plot of $\Delta F_c$ and
$\Delta F_l$ is shown in Fig.~\ref{deltaF}. It is clear that
$\Delta F_l$ is nearly constant as a function of $\Omega$ while
$\Delta F_c$ goes through a minimum in the range of $\Omega =
0.065$ to 0.12. A simple order of magnitude calculation gives the
Kramers rate \cite{kramers} $T_f^{-1} =\frac{(F^{\prime \prime}_{min} \vert
F^{\prime \prime}_{saddle} \vert )^{1/2}}{2\pi} exp -(\Delta F/D)
\sim  10^{-5}$ which matches with the frequency range  of the
first harmonic seen in the PSD (see Fig.~\ref{power}). We note
here that the first passage time for the system to cross the
barrier from the liquid state is much too large as the barrier
$\Delta F_l \sim 0.01$.
\begin{figure}
\epsfig{figure=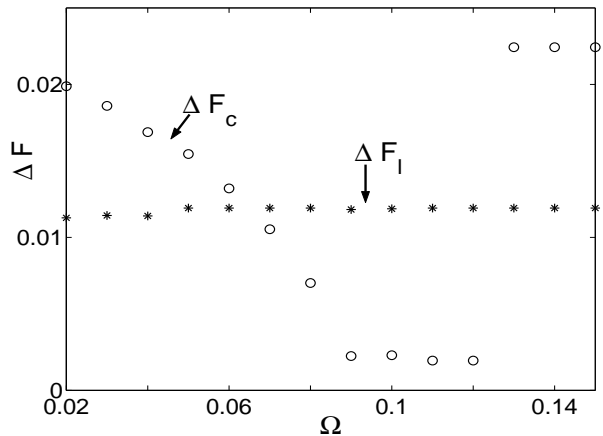,width=8cm,height=6cm}
\caption{$\Delta F_c$ and $\Delta F_l$ as a function of $\Omega$.}
\label{deltaF}
\end{figure}          
Thus, it is clear that the  reduced model reproduces,
qualitatively, most features  of the particle model.  The time
series of $S({\bf k},t)$ in the particle model is very similar to
that of $\rho(t)$ which we have demonstrated has all the features
of stochastic resonance.  The regime showing stochastic resonance
is more pronounced for optimum values of $\Omega$ and $\alpha$. In
fact here, as in our particle model simulations, in the melt-freeze
regime, for low values of the coupling parameter, the crystalline
state is favored and for high values of the coupling parameter, the 
liquid state is favored, while in an intermediate regime they enter
roughly equally. The similarity of the reduced model with the particle 
model is well summarized by the phase diagram of the model in the 
$\Omega-\alpha$ plane which is similar to that of the particle model in 
the $F^*_d - \epsilon$ plane.

\section{Discussion}

In summary, we have studied the nonequilibrium statistical behavior of 
two adjacent monolayers sheared past each other, using
Brownian dynamics simulations.  For low and high driving forces,
we obtain macroscopically ordered, steadily drifting states. In a
suitable range of driving rates we see that the system
switches between crystalline and liquid like states. The
residence times are nearly equal  in a intermediate range of
values of the inter-layer coupling. As we have seen, the
interlayer coupling essentially determines the barrier each
particle has to surmount in order keep pace with the applied
drive. Our simulations show
that the switching between the crystalline and liquid like states
maintains the spatial coherence ( or the lack of it in the liquid
state) and thus is an example of {\em cooperative stochastic resonance}.
Although there is no external imposed time-periodic potential
present in our system, its role is played by  the drive which
shears the two layers past each other. Thus, each moving layer provides a
time-varying potential.  To make the connection to SR more
concrete, we have introduced a reduced model for the dynamics of amplitude
and phase of the crystalline order parameter \cite{rlsrsmelt} which mimics all
features of the particle model apart from showing the main feature
of SR namely the enhancement of signal to noise ratio for 
optimum values of noise strength. This relation between the particle 
model and the reduced model makes a convincing case that the 
melt-freeze cycles observed in the former are indeed a manifestation 
of stochastic resonance of a spatially extended noisy interacting system 
subjected to a constant drive.  

A few comments may be in order on the dynamics of the particle
model. Although our model falls broadly into the class of  shear 
driven systems which are known to display stick-slip behavior \cite{homola,
thompson,vinokur} 
in most such cases the system spends considerable time in the stuck phase 
and very little time in the slip phase. Experiments on confined fluids of few
monolayers (a situation  superficially similar to our model) and
the corresponding  molecular dynamics simulations
\cite{homola,thompson} show that the system also displays melt-freeze
cycles  with the crystalline phase persisting significantly much
longer than the melt phase. From this point of view, the persistence dynamics 
we observe with nearly equal residence times in the crystalline state and the
liquid-like states comes as a surprise and is clearly not conventional 
stick-slip. The reduced model has helped us to elucidate the connection to
stochastic resonance apart from the similarity  of the phase
diagrams of the two models. However, it is clear that not all
aspects of the particle model are mimicked by the reduced
model. For instance, the feature observed for small interlayer
coupling $\epsilon$, namely the smaller extent of order as  seen
by the small values of $S({\bf k},t)$ in the crystalline state (
compared to larger values of $\epsilon$) and larger fluctuations
is not seen in the reduced model. As explained earlier, this
feature is a many body effect appearing in a low barrier
situation and therefore cannot be explained  on the basis of the
SR features of the reduced model valid only in the Kramers high
barrier limit (even if one were to include appropriate spatial
degrees of freedom).

Finally, let us comment on the experiments which can verify our
simulations. We expect that this phenomenon should arise in
adjacent crystal planes of sheared colloidal crystals. To study this effect in
bulk sheared colloidal crystals would require not conventional scattering probes but something that focuses on an adjacent pair of crystal planes \cite{grier}.
Alternately we could look at two solid surfaces patterned with ordered 
copolymer \cite{copoly1,copoly2} or colloidal monolayers under relative shear. The
colloids or copolymer adsorbed to the two surfaces must be of two
different kinds, each having more affinity towards one of the
surfaces. Another  possibility would be to study electrokinetic
motion of charged 2d confined binary colloids in the presence of a constant
external electric field, generalizing the ideas of \cite{lowen}.
Centre of mass measurements would be the same as that made for two
species being driven in opposite directions.

\acknowledgments

MD thanks CSIR, India for financial support, SERC, IISc for providing
computational facilities and C. Dasgupta, B.  Chakrabarti and C. Das for 
useful discussions.


\begin{thebibliography}{000}
\bibitem{v1} The local variance of the velocity is not, of course, a good 
measure of long-range correlations. To detect the latter we would need to 
measure the spatial correlation function of the drift velocity, which we 
have not done.
\bibitem{f1} Note that the equations of motion do not have a factor $\rho$
multiplying $\dot\theta$, so that even though $\theta$ is periodic, we have a
planar metric.
\end{thebibliography}
\end{document}